\documentclass[conference]{IEEEtran}
\usepackage{balance}
\IEEEoverridecommandlockouts

\usepackage{amsmath,amssymb,amsfonts}
\usepackage{cleveref}
\usepackage{algorithmic}
\usepackage{graphicx}
\usepackage{caption}
\usepackage{subcaption}
\usepackage{textcomp}
\usepackage{xcolor}
\usepackage{changepage}
\usepackage{url}
\usepackage[numbers]{natbib}
\usepackage{atbegshi}

\AtBeginDocument{\AtBeginShipoutNext{\AtBeginShipoutDiscard}}

\def\BibTeX{{\rm B\kern-.05em{\sc i\kern-.025em b}\kern-.08em
    T\kern-.1667em\lower.7ex\hbox{E}\kern-.125emX}}

\crefformat{equation}{equation~(#2#1#3)}
\Crefformat{equation}{Equation~(#2#1#3)}    
    
\begin{document}

\title{Incentivizing Private Data Sharing in Vehicular Networks: A Game-Theoretic Approach}

\thanks{}
\author{
    \IEEEauthorblockN{Yousef AlSaqabi\IEEEauthorrefmark{1}\IEEEauthorrefmark{2}, Bhaskar Krishnamachari\IEEEauthorrefmark{1}}
    \IEEEauthorblockA{\IEEEauthorrefmark{1}University of Southern California, USA}
    \IEEEauthorblockA{\IEEEauthorrefmark{2}Kuwait University, Kuwait}
    \{alsaqabi, bkrishna\}@usc.edu
}

\maketitle

\begin{abstract}  

In the context of evolving smart cities and autonomous transportation systems, Vehicular Ad-hoc Networks (VANETs) and the Internet of Vehicles (IoV) are growing in significance. Vehicles are becoming more than just a means of transportation; they are collecting, processing, and transmitting massive amounts of data to make driving safer and more convenient. However, this advancement ushers in complex issues concerning the centralized structure of traditional vehicular networks and the privacy and security concerns around vehicular data. This paper offers a novel, game-theoretic network architecture to address these challenges. Our approach decentralizes data collection through distributed servers across the network, aggregating vehicular data into spatio-temporal maps via secure multi-party computation (SMPC). This strategy effectively reduces the chances of adversaries reconstructing a vehicle's complete path, increasing privacy. We also introduce an economic model grounded in game theory that incentivizes vehicle owners to participate in the network, balancing the owners' privacy concerns with the monetary benefits of data sharing. This model aims to maximize the data consumer's utility from the gathered sensor data by determining the most suitable payment to participating vehicles, the frequency in which these vehicles share their data, and the total number of servers in the network. We explore the interdependencies among these parameters and present our findings accordingly. To define meaningful utility and loss functions for our study, we utilize a real dataset of vehicular movement traces.

\end{abstract}


\section{Introduction}

As we move towards a future where smart cities and autonomous transportation systems are gaining momentum, the role of Vehicular Ad-hoc Networks (VANETs) and the Internet of Vehicles (IoV) becomes increasingly significant~\cite{IoV}. In this rapidly evolving landscape, vehicles are evolving from being a means of transportation into key parts of a large, interconnected network~\cite{Hao}. Entities like traffic management systems, autonomous vehicles, and infotainment service providers, known as data consumers, use this shared data to improve their services and operations. For example, traffic management systems can use real-time data to manage congestion, enhance route planning, and decrease travel times~\cite{Soto2021ASO}. Autonomous vehicles can leverage this shared data for improved decision-making, leading to safer and more efficient operations~\cite{ameen2020review}. Infotainment services can offer personalized content based on user preferences and location data~\cite{cellular_v2x}. Looking ahead, private data sharing in VANETs is set to offer a growing number of applications. However, this shift in the transportation landscape also brings with it a variety of complex challenges that require careful examination and innovative solutions.

The centralized nature of traditional vehicular network architecture, in particular, poses a significant concern. Central servers bear the brunt of increasing traffic loads and any failure can lead to substantial disruption~\cite{security_challenges_iov}, making the system vulnerable. This necessitates a shift towards decentralized network management and distributed storage solutions for a more resilient, scalable, and efficient IoV~\cite{jiang_blockchain-based_2019}.

Simultaneously, the increasing network complexity and the sensitive nature of vehicular data raise critical questions about privacy and security~\citep{security_challenges_iov, Ogonji, chen2022}. The indiscriminate sharing of vehicular data, such as vehicle locations and sensor readings, could lead to severe privacy violations and potential misuse~\cite{chen2022, garg2020survey}. Despite the growing body of research focused on secure communication and privacy protection in vehicular networks~\cite{security_challenges_iov, garg2020survey, bc_iov_survey, Blockchain_solutions_IoV, blockchain_security_services}, there remain critical gaps in addressing these concerns while simultaneously incentivizing users to participate in data sharing. 

This paper proposes a novel, game-theoretic network architecture to address these challenges. Our approach decentralizes the data collection process by introducing distributed servers across the network. Vehicles would periodically send their data to one of these servers at random, where the servers would then aggregate the data into spatio-temporal maps using secure multi-party computation (SMPC)~\cite{SMPC}. The aggregated data would then be forwarded to a central server for further analysis by the data consumer. This scheme effectively reduces the likelihood of an adversary being able to reconstruct the complete path of a vehicle, even if they manage to intercept some of the individual data samples.

The application of this approach involves an economic model inspired by the Stackelberg competition~\cite{gametheory}, a concept in game theory, wherein the data consumer and the vehicle owners are the key players. In this game, the data consumer assumes the role of the leader, making the first move by defining the number of servers in the network, the compensation provided to the participating vehicles, and the frequency with which these vehicles share their data. The vehicle owners follow by evaluating the terms set by the data consumer, and decide whether or not to participate in the data provision based on their privacy concerns and potential monetary gains from data sharing. This model ensures that the servers' operational costs are managed and also aims to maximize the data consumer's utility from the gathered sensor data. 

We also propose that the agreements between the vehicles, servers, and the data consumer be made autonomous, transparent, and trustworthy through the use of blockchain-based smart contracts. These contracts encourage participation and maintain the system's integrity by tracking individual contributions and handling payments according to a predefined structure~\cite{smartcontracts}. In addition, they also enforce an audit and verification process to prevent false data submissions, thereby ensuring the reliability and credibility of the shared data~\cite{blockchain_security_services}.

The remainder of this paper is organized as follows: Section~\ref{related} provides a review of the related work. Section~\ref{formulation} outlines our proposed network. Section~\ref{modelling} explains the economic modelling equations used. Section~\ref{results} presents the results of our optimization and sensitivity analysis. Finally, Section~\ref{conclusion} concludes the paper and proposes directions for future research.

\section{Related Work} \label{related} 

\subsection{Private Data Sharing}

The concept of data sharing in vehicular networks has been explored in various studies, each proposing unique solutions to address the challenges of privacy, security, and efficiency.

Several studies have proposed different methods to enhance data privacy and security in vehicular networks. Kaiser et al.~\cite{Kaiser} proposed an Open Vehicle Data Platform that uses Blockchain technology to ensure the privacy of vehicle owners and drivers during the exchange of vehicle and driving data. Kong et al.~\cite{kong_privacy-preserving_2019} proposed an efficient and location privacy-preserving sensory data sharing scheme with collision resistance in IoV. Their scheme uses the modified Paillier Cryptosystem to achieve location privacy-preserving multi-dimensional sensory data aggregation. Fan et al.~\cite{fan_cloud-based_2021} proposed a cloud-based mutual authentication protocol aiming at ensuring efficient privacy preserving in IoV system.


A few studies have proposed incentive mechanisms to encourage data sharing. Zhang and Xu~\cite{zhang_blockchain-based_2022} proposed a mechanism that combines certificateless message authentication and blockchain incentives to provide anonymity and non-repudiation for traffic-related message reporters. Yeh et al.~\cite{yeh_blockchain-based_2022} proposed a fair and privacy-preserving data query system for vehicle networks based on blockchain technology. Their scheme provides features of sustainable data accessibility and a large amount of data storage and also provides an incentive mechanism to encourage users to share their traffic information.

\subsection{Game Theory}

Game theory, though not extensively prevalent, has found niche applications within the realm of vehicular networks, catering to specific use cases and scenarios. One such application is in the field of data security; Gupta et al.~\cite{gupta_game_2022} proposed a blockchain-enabled game theory-based authentication mechanism to secure Internet of Vehicles (IoV) settings.

Another application of game theory in vehicular networks is in the area of computation offloading. Liwang et al.~\cite{liwang_game_2019} developed an opportunistic Vehicle-to-Vehicle (V2V) computation offloading scheme based on game theory. They formulated the computation offloading scheme and pricing strategy as a Stackelberg game, taking into account various factors such as vehicular mobility models, V2V contact durations, computational capabilities, channel conditions, and service costs. Similarly, Hassija et al.~\cite{hassija_dagiov_2020} used a game theoretic approach to map service providers and consumers for performing offloading services in a cost-optimal way in V2V communication.


In the realm of vehicular networks, the application of game theory has been primarily focused on areas such as data security, computation offloading, and resource allocation. However, our research diverges from these established paths by leveraging the Stackelberg game theory to incentivize private data sharing within the network. While previous studies, such as that of Liwang et al.~\cite{liwang_game_2019}, have also utilized the Stackelberg game, our work uniquely applies it to the challenge of balancing data sharing incentives with privacy preservation. Furthermore, the introduction of monetary incentives in our model represents a novel approach in the field, distinguishing our research from existing literature.

\subsection{Secure Multi-Party Computation}
Secure Multi-Party Computation (SMPC) is a protocol that enables distributed parties to jointly compute an arbitrary functionality without revealing their own private inputs and outputs~\cite{SMPC}. SMPC has been used for various security applications, including privacy-preserving machine learning with multiple sources~\cite{bost2014machine}, private set operations~\cite{Freedman}, and genome sequence comparison for secure genomic computation~\cite{genome}. SMPC is commonly constructed with cryptographic schemes such as homomorphic encryption, which is an encryption technique that allows applying computations on the encrypted data without having to decrypt it~\cite{homomorphic}. 

In the realm of vehicular networks, Song et al.~\cite{SMPC_Vanet} proposed an anonymous authentication scheme based on SMPC to solve securrity and privacy problems for VANETs. Raja et al.~\cite{AI_Blockchain_IoV} combined blockchain technology with AI to overcome security challenges in IoV, speed up transaction verification, and optimizes energy consumption. The authors of~\cite{AutoMPC} proposed a cooperative control strategy involving SMPC that performs computations while increasing resillience towards latency and adversaries. Gupta et al.~\cite{HEncryption} have combined SMPC with homomorphic encryption and proposed a scheme that provides real-time location tracking using GPS without divulging the actual location of the user.

Our research builds upon these existing works by proposing the use of SMPC for secure data aggregation between servers within vehicular networks. While previous studies have focused on applications such as anonymous authentication and location tracking, our work introduces a novel application of SMPC for preserving privacy during data aggregation. Therefore, our research not only extends the application of SMPC in vehicular networks but also contributes a unique solution to the challenge of privacy-preserving data aggregation.



\section{Problem Formulation} \label{formulation}

\begin{figure}[!t]
    \centering
    \includegraphics[scale=0.4]{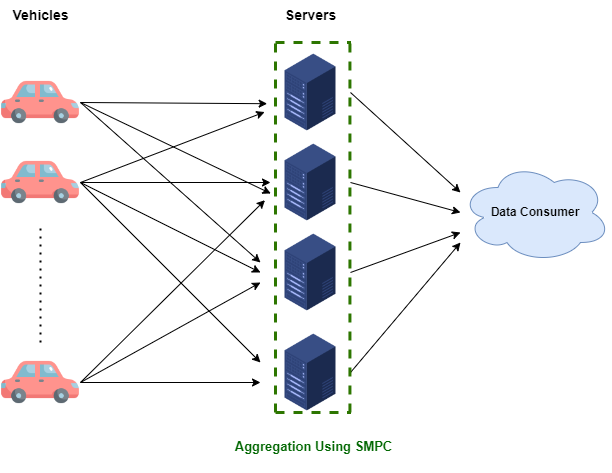}
    \caption{Network Flow Diagram}
    \label{fig:diagram}
\end{figure}

Our network architecture is made up of three key components: the vehicles providing the data, the independent servers processing the data, and the data consumer collecting the processed data. As depicted in Fig~\ref{fig:diagram}, vehicles periodically transmit their sensor data, which could include location coordinates or specific measurement readings, to one randomly selected server at a time. This approach ensures that no single server has access to the full raw data of any one vehicle, enhancing the privacy and security of the data.

The servers employ Secure Multi-Party Computation (SMPC) to aggregate the sensor data into spatio-temporal maps. Generally speaking, SMPC is a protocol that allows multiple parties to jointly compute a function over their inputs while keeping those inputs private. In our context, it ensures that the raw sensor data from the vehicles sent to one server is not exposed to any other servers or external entities.

Once the data is aggregated into spatio-temporal maps on each server, it is forwarded to a central server for further analysis. This process allows us to extract valuable insights from the data while preserving the privacy of the individual vehicles' raw data.

\subsection{Vehicles}

The vehicles share private, signed information such as location data to a random receiving server. In exchange, the vehicles are compensated based on the amount of data they transmit.  However, the higher the volume of data transmitted, the greater the privacy risk, given that the information sent includes time-stamped location data.

We model the utility of the vehicle by:

\begin{equation} \label{vutil}
    c_1\cdot f_d - L(f_d,s)
    \end{equation}

where $c_1$ is the payment provided to the vehicle, $f_d$ is the frequency of data transmission, and $s$ is the number of servers. We assume that all vehicles share data at the same frequency $f_d$. $L(f_d,s)$ is a function that says the privacy loss increases with $f_d$ and decreases with $s$. This loss function will be explained in greater detail in Section~\ref{4b}.

Because of this trade-off, not every vehicle would be willing to contribute its data to the network; the utility gained would have to exceed the perceived privacy loss of sharing its data for the vehicle to have positive utility and therefore agree to contribute its data.  

This will create a supply curve because as $c_1$ increases, more vehicles would be willing to provide data for a given $f_d$. Let's model a function $v(c_1, f_d, s, V)$ as the number of vehicles that provide data to the network when the vehicles are paid $c_1$, each vehicle sends data at a frequency of $f_d$, there are $s$ servers, and a maximum of $V$ vehicles registered in the system. 

To find $v(c_1, f_d, s, V)$, we consider that the privacy sensitivity of each individual owning a vehicle can be modeled by a random variable $e_i$, with a given cdf $F_{e_i}(\cdot)$. For an individual with sensitivity $e_i$, the utility due to privacy is given as: 

\begin{equation} \label{vutil2}
    w_i = c_1\cdot f_d - e_i\cdot L(f_d,s)
    \end{equation}

If $w_i > 0$, the individual would provide their data. The probability that this happens depends on $e_i$.  Using \cref{vutil2}, we have that for positive $w_i$ values, the probability of an individual agreeing to share their data would be:

\begin{equation} \label{vutil3}
  Pr(w_i > 0) = Pr\left(e_i < \frac{c_1 \cdot f_d}{L(f_d,s)}\right)  = F_{e_i}\left(\frac{c_1 \cdot f_d}{L(f_d,s)}\right)
    \end{equation}

For our simulations, we use a log-normal distribution for $F_{e_i}(\cdot)$. If we have $V$ total vehicles in the system, the expected number of vehicles that would participate in the network would be: 

\begin{equation} \label{vutil4}
    v(c_1, f_d, s) = V\cdot F_{e_i}\left(\frac{c_1 \cdot f_d}{L(f_d,s)}\right)
    \end{equation}
    
\vspace{5pt}

\begin{equation} \small \label{vutil5}
    = V\cdot\frac{1}{\sqrt{2\cdot\pi \cdot \sigma ^2}}  \cdot \frac{1}{\frac{c_1 \cdot f_d}{L(f_d,s)}} \cdot \exp\left[-\frac{\left(\ln\left(\frac{c_1 \cdot f_d}{L(f_d,s)}\right) - \mu\right)^2}{2\sigma^2}\right]
    \end{equation}

\vspace{5pt}

\subsection{Servers}

We assume that each server in our network is responsible for collecting and processing the data shared by the vehicles. Each server incurs a cost proportional to the amount of data it processes, in addition to a fixed maintenance or upkeep cost. In our proposed framework,  one or more third parties could be in charge of server management. This keeps the data consumer separate from server administration, enhancing user privacy and trust. By distributing control, we make sure no single entity can access all the raw data, reinforcing our dedication to data privacy. The key assumption we make, however, is that the data consumer shoulders the cost for the use of these servers. The total cost incurred per server would be:

\begin{equation} \label{servercost}
    c_2 \cdot v(c_1, f_d, s, V) \cdot \frac{f_d}{s} +  c_3
    \end{equation}

where $c_2$ and $c_3$ are the server computation and upkeep costs respectively. 

\subsection{Data Consumer}

The data consumer is the entity that provides the payments to the vehicles and servers and receives the final aggregated data. The data consumer's total utility is therefore modeled as:

\begin{equation} \label{profit}
\begin{aligned}
&\quad U(v(c_1,f_d,s), f_d)\\ 
&- (c_2\cdot v(c_1,f_d,s, V)\cdot\frac{f_d}{s} + c_3)\\ 
&- c_1\cdot v(c_1,f_d,s, V)\cdot f_d
\end{aligned}
\end{equation}

\Cref{profit} represents the utility gained from the vehicles' data minus the total cost to run the servers and the total payment made to the participating vehicles. This is the profit equation that we will optimize in this paper. The data consumer determines the compensation allocated to the participating vehicles and servers, as well as the total number of servers in the network. Once the data consumer establishes the servers' payments ($c_2$ and $c_3$), the optimization process delivers the optimal values for vehicle compensation ($c_1$), server count ($s$), and data transmission frequency ($f_d$) that maximize the data consumer's overall utility.

\vspace{5pt}

\section{Economic Modelling} \label{modelling}

\subsection{Data Consumer's Utility Function}


The data consumer's utility gained from vehicle data can be expressed as $U(v(c_1,f_d,s), f_d)$. It depends on the number of vehicles providing data, which in turn depends on the payment made to each vehicle for providing their data $c_1$, the frequency in which these vehicles are sharing their data $f_d$, and the total number of servers in the network $s$. 

We model the data consumer's utility in a given area as an increasing function related to the number of vehicles contributing data, with diminishing utility as that number increases: 

\begin{equation} \label{util}
    -1/( 1 + a\cdot exp( \frac{-1}{\sqrt{n}}))+1
    \end{equation}

In \cref{util}, $n$ is the number of vehicles contributing data, and $a$ is a variable parameter determining the limit to which the utility function converges. When $a$ increases, this convergence tends towards 1. The specific value of $a$ is less significant as long as it is sufficiently large. If $a$ becomes too small, the function will converge to a lesser value. For our demonstration, we choose $a$ to be equal to 100. 

To model this function, we used a dataset containing timestamped coordinates of 2927 taxis in Beijing over a 24-hour time period\footnote{This dataset was obtained from University of Southern California’s Autonomous Networks Research Group (https://anrg.usc.edu/www/downloads/)}. Since the dataset provides data points for every minute, the $f_d$ term will be described in units of samples per minute.

\begin{figure}[!t]
    \centering
    \includegraphics[scale=0.65]{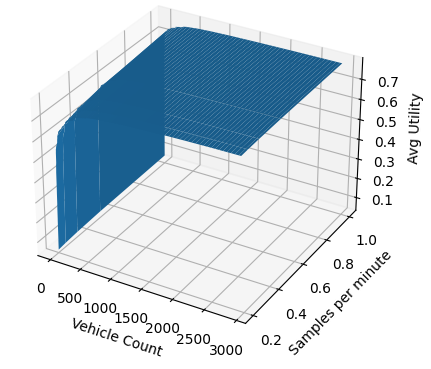}
    \caption{Plot of data consumer's average utility as a function of the number of vehicles and the frequency of data transmission}
    \label{fig:avgutil}
\end{figure}

Our objective is to discern the data consumer's average utility if they had a sub-sample of this dataset. We begin by sectioning the entire dataset into spatio-temporal grids. Within each grid, we use the number of vehicle samples along with \cref{util} to calculate the utility. We then compute the average utility across all grids to determine the average utility of the dataset's region. We repeat this process with various $V$ and $f_d$ values to create a 3D plot as shown in Fig.~\ref{fig:avgutil}. This plot illustrates how the average utility for the data consumer changes as the number of vehicles and the frequency of data transmission increase. Next, we use the curve fitting function shown in \cref{uvfd} to establish an equation that expresses the utility gained from the vehicles' data $U(v(c_1,f_d,s), f_d)$.

\begin{equation} \label{uvfd}
    U(v(c_1,f_d,s), f_d) = \alpha \cdot(1-\exp(-\beta \cdot(v \cdot f_d)))
    \end{equation}


\vspace{5pt}

After curve fitting, our $\alpha$ and $\beta$ constants came out to be 0.99 and 0.45 respectively. We show how sensitive our results are to these parameters in Section~\ref{results}.

\subsection{Loss Function} \label{4b}
The vehicle's privacy loss function, denoted as $L(f_d,s)$, depends on the frequency in which the vehicle is sharing its data and the number of servers it is sharing its data to. We provide a model to quantify this loss in terms of how accurately a potential adversary can reconstruct a vehicle's driving path given a certain number of samples. Given a set of locations on a vehicle's path, an adversary's best estimate of the path traversed by the vehicle would be the shortest path routes between each pair of those points. Our measure of privacy would be how different this reconstructed path would be compared to the vehicle's original path. 

\begin{figure}[!t]
    \centering
    \includegraphics[scale = 0.3]{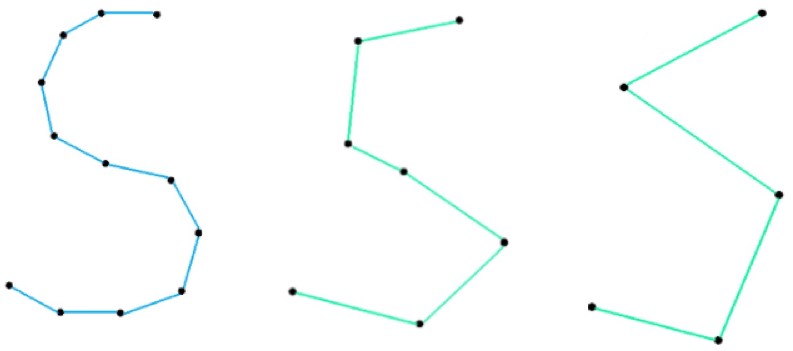}
    \caption{Paths with Different Number of Data Points}
    \label{fig:trips}
\end{figure}

Figure~\ref{fig:trips} presents an example of an original trip represented in blue, along with two sub-trips, shown in green, that contain a subset of the data points from the original trip. It is evident that paths are less similar when fewer data points are available. Consequently, paths with less similarity offer greater privacy to the user.

First, we find how much privacy is lost to one server that receives $f_d/s$ samples per minute from a given vehicle. To do this, we measure our privacy metric empirically using our dataset. We sample one trip from the dataset, which will have a $f_d$ of one sample per minute, and compare that same trip to subtrips with less frequent samples, ranging from $f_d = 1/2$ to $f_d=1/10$. We use the similarity library within the TensorBay package~\cite{tensorbay} to determine the path similarity between our initial trip and all subtrips. This library utilizes the Fr\'echet distance~\cite{frechet} as a measure of comparing two curves. Finally, we repeat this process for all the vehicles in the dataset to find the average similarity scores as the value of $f_d$ changes.  

\begin{figure}[!h]
    \centering
    \includegraphics[scale=0.22]{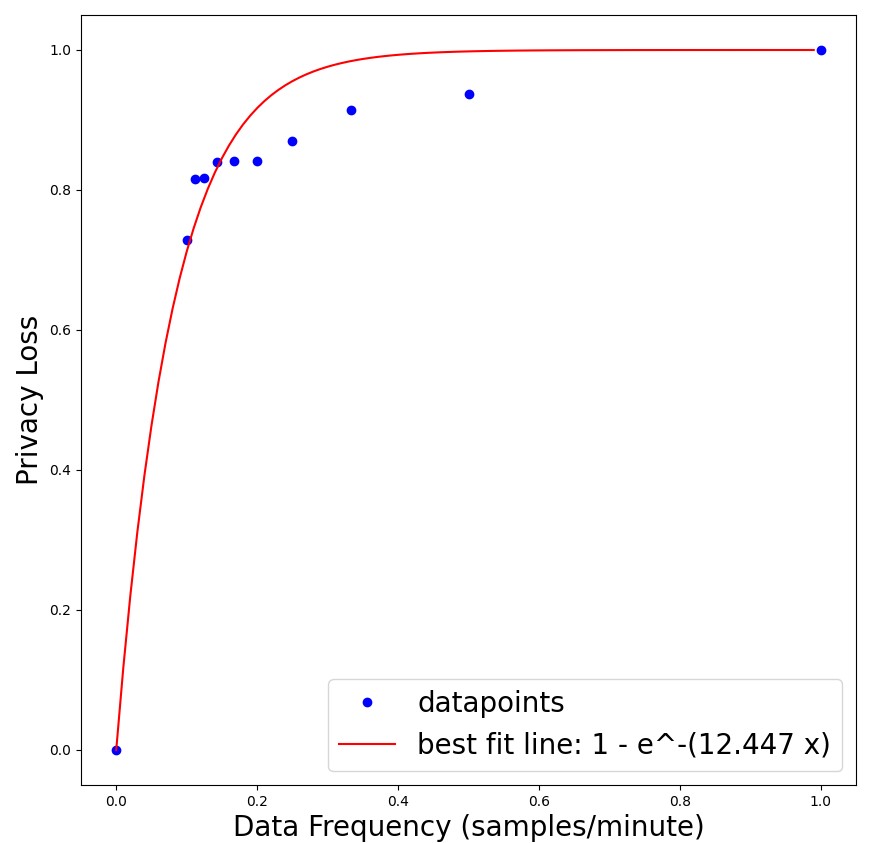}
    \caption{Privacy Loss as a function of the frequency of data transmission}
    \label{fig:similarity}
\end{figure}

Fig~\ref{fig:similarity} shows that when the frequency of data sharing increases, it becomes easier for a potential adversary to reconstruct a user's path, leading to higher privacy loss. We can also see the curve-fitted equation that gives us the privacy loss as a function of $\frac{f_d}{s}$. The loss function takes values from 0 to 1, with 1 being 100\% privacy loss. However, the network's privacy is also influenced by the total number of servers and the frequency of data transmission as well. The total loss equation is therefore:

\begin{equation} \small \label{loss}
\begin{split}
    L(f_d,s) = & 1 - \exp\left(-12.447 \cdot \frac{\mathit{f_d}}{s}\right) \\
               & - \exp\left(-p \cdot \mathit{f_d}\right) \\
               & - \exp\left(-\frac{q}{s}\right)
\end{split}
\end{equation}

\Cref{loss} contains three terms: privacy loss from $\frac{f_d}{s}$, $f_d$, and $s$. Each term independently influences the overall loss function. The first term represents the privacy loss per server, the second term indicates that the loss function approaches 1 as $f_d$ approaches infinity, and the third term confirms that the loss function tends to zero when $s$ approaches infinity. The constants $p$ and $q$ are selected to ensure that both the $f_d$ and $s$ variables contribute adequately. Although it would have been possible to set both constants as equal, we opted to weigh the influence of adding another server to the privacy loss more heavily than that of adding another sample per minute, in the context of our optimization. In demonstrating this, we have set $p$ and $q$ at 0.1 and 10 respectively.

\section{Results} \label{results}
\subsection{Optimization}

The optimization problem for the data consumer involves selecting the optimal values for the number of servers ($s$), frequency of data transmission ($f_d$), and the payment provided to the vehicle ($c_1$) in order to maximize the total profit, as expressed in \cref{profit}. Each of these three parameters is crucial to optimize, as they each involve specific trade-offs. Increasing the number of servers enhances overall privacy but also raises the network's operating costs. A higher $f_d$ yields greater utility for the data consumer, but may deter vehicles from sharing their data due to privacy concerns. Lastly, raising $c_1$ could attract more vehicles to participate in the network, thus increasing utility, but this also incurs higher costs for the data consumer.

\begin{figure*}[ht]
    \centering
    \includegraphics[scale=0.36]{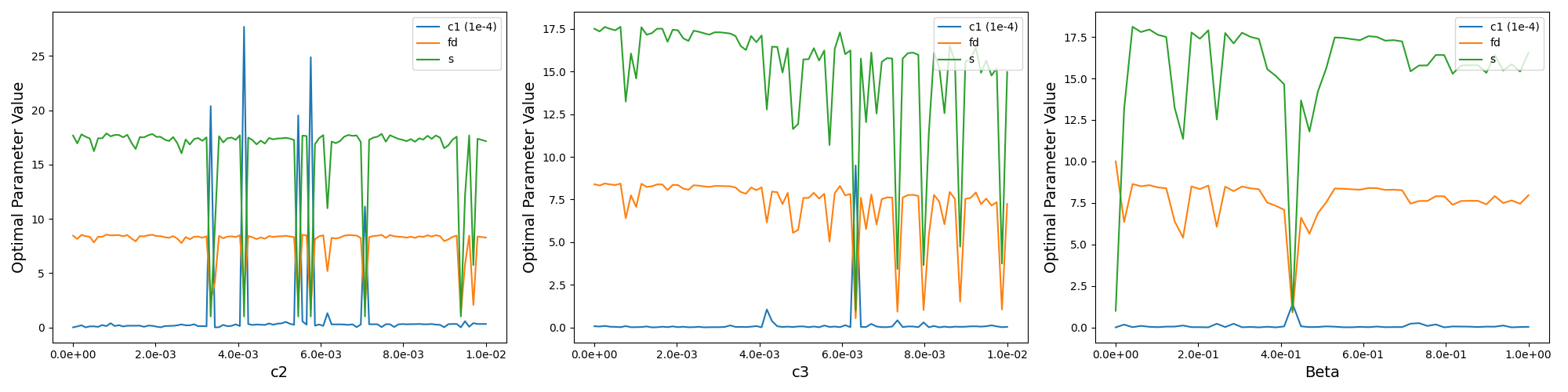}
    \caption{Sensitivity Analysis for $c2$, $c3$, and $\beta$ respectively.}
    \label{fig:sens_analysis}
\end{figure*}

To optimize \cref{profit}, we must first choose acceptable values for the server computation cost ($c_2$), the server upkeep cost ($c_3$), the total number of vehicles $V$, and the mean and variance of our log-normal distribution function. We set the number of total vehicles to be 2928, which is the number of independent vehicles in our dataset. We use a mean of 0 and standard deviation of 0.5 for our log-normal distribution to have a positive-shaped bell curve.

The following set of equations guides our choices for $c_2$ and $c_3$. As depicted in \cref{opt1}, $c_3$ should be on the order of the payment distributed to each vehicle per server. \Cref{opt2} emphasizes that $c_3$ must stay below the computational expense the server bears for each vehicle that shares data. Lastly, \cref{opt3} suggests that $c_2$ should be less than the reciprocal of the total number of vehicles in the network. It is important for our chosen values to comply with these equations, as the relative values of the parameters are more significant than their absolute values. For this reason, we choose 1e-6 and 1e-4 for $c_2$ and $c_3$ respectively and perform sensitivity analysis on these constants in the subsequent subsection.

\begin{subequations} \label{opt}
\begin{alignat}{2}
    &c_3 &&\sim c_1 \cdot \frac{V}{s} \label{opt1}\\
    &c_3 &&< c_2 \cdot \frac{V}{s} \label{opt2} \\
    &c_2 &&< \frac{1}{V} \label{opt3}
\end{alignat}
\end{subequations}

\vspace{5pt}

We used the scipy.optimize package~\cite{2020SciPy-NMeth} to solve our optimization problem. We observe from Table~\ref{OptResults} that the optimized values are 3.57e-6 paid to the vehicle, 7.31 samples per minute shared to the servers by the vehicles, and 15.12 total servers.


\begin{table}[ht] 
  \begin{adjustwidth}{-0.5\columnwidth}{-0.5\columnwidth}
    \centering 
    \captionsetup{font={footnotesize}} 
    \caption{Optimization Results}
    \label{OptResults}
    \resizebox{0.4\textwidth}{!}{
      \begin{tabular}{|c|c|c|c|}
        \hline
        Parameter & $c_1$ & $f_d$ (samples/min) & servers \\
        \hline
        Optimized Value & 3.57e-6 & 7.31 & 15.12 \\
        \hline
      \end{tabular}
    }
  \end{adjustwidth}
\end{table}

\subsection{Sensitivity Analysis} \label{sensitivity}

In this section, we discuss the sensitivity analysis performed to determine the effects of varying specific constants on our optimization results. The variables analyzed include the server computation cost ($c_2$), server upkeep cost ($c_3$), total number of vehicles ($V$), standard deviation of individuals' privacy sensitivity, and the constant $\beta$ found in \cref{uvfd}. We're particularly interested in how these changes influence the three optimized parameters: the vehicle payment ($c_1$), frequency of data transmission ($f_d$), and the number of servers ($s$).

Altering the total number of vehicles ($V$) does not considerably affect $c_1$, $f_d$, or $s$. However, it impacts the number of vehicles participating in the network, as these are a subset of the total vehicles. Therefore, an increase in the total number of vehicles leads to more vehicles participating in the network.

When we increase the standard deviation of individuals' privacy sensitivity, we observe a slight increase in the number of servers required, a slight increase in the frequency of data sharing, and a decrease in the vehicle payment. This result aligns with the understanding that a larger variability in privacy sensitivity necessitates more servers and reduced data sharing frequency to enhance privacy. Consequently, the vehicles demand less payment due to the lowered privacy risk posed by the network.

Figure~\ref{fig:sens_analysis} illustrates the sensitivity of our results when parameters $c2$, $c3$, and $\beta$ are adjusted. This figure shows that the various parameters are largely robust to values of $c_2$, $c_3$, and $\beta$, with the exception of particular values that show sharp changes that could be attributed to numerical instability of the non-linear optimization. However, even where these sharp changes occur, we observe that $f_d$ and $s$ exhibit the same trends, increasing and decreasing in unison. Conversely, $c1$ behaves differently, showing a sharp increase whenever $f_d$ and $s$ experience drops. This particular surge is in line with expectations since the model needs to be balanced due to sudden drops in $f_d$ and $s$.  The key insight from these graphs is that the absolute values of these parameters are less significant than their values in relation to the other parameters.

\section{Conclusion} \label{conclusion}

In this paper, we introduced a novel, game-theoretic, privacy-preserving network architecture that incentivizes private data sharing in VANETs. This approach decentralizes data collection and integrates distributed servers across the network, where vehicles would periodically share their data with random servers that aggregate the data into spatio-temporal maps using secure multi-party computation (SMPC). This model increases user privacy by denying potential adversaries the ability to reconstruct the complete path of a vehicle, even if they manage to intercept some of the individual data samples. This approach also introduces an economic model that balances the drivers' privacy concerns with the monetary benefits of data sharing, while managing the operational costs of the servers in the network. The model optimizes the overall utility for the data consumer by determining the most suitable payment to vehicles ($c_1$), frequency of data sharing ($f_d$), and the number of servers ($s$) in the network. 

For future development, we aim to improve our optimizer by employing a hybrid of Neural Networks and Genetic Algorithms, enhancing optimization stability and efficiency. Concurrently, we plan to integrate tailored blockchain solutions for our network through meticulous requirement analysis, hybrid blockchain adoption, and smart contract design. These advancements seek to provide novel insights and bolster security, trust, operational efficiency, and user empowerment in IoV applications.
\newpage


\bibliographystyle{unsrt}
\bibliography{ref}

\end{document}